\documentclass[conference]{IEEEtran}
\pdfoutput=1
\usepackage[hyphens]{url}
\usepackage{hyperref}
\hypersetup{colorlinks,allcolors=blue}
\usepackage[backend=biber, style=ieee]{biblatex}
\usepackage{xurl}

\usepackage{amsmath,amssymb,amsfonts}
\usepackage{algorithmic}
\usepackage{graphicx}
\usepackage{textcomp}
\usepackage[table,xcdraw]{xcolor}
\usepackage{balance}
\usepackage{multirow}
\usepackage{amsmath}
\usepackage{array}

\usepackage[table,xcdraw]{xcolor}
\usepackage{balance}
\usepackage[flushleft]{threeparttable} 
\usepackage{soul} 
\usepackage{orcidlink}
\usepackage{makecell}

\def\BibTeX{{\rm B\kern-.05em{\sc i\kern-.025em b}\kern-.08em
    T\kern-.1667em\lower.7ex\hbox{E}\kern-.125emX}}

\usepackage{geometry}
\IEEEoverridecommandlockouts
\begin{document}

\title{Bi-Directional Transformers vs. word2vec: Discovering Vulnerabilities in Lifted Compiled Code}

\author{Gary A. McCully\,\orcidlink{0009-0005-6163-7720}\textsuperscript{*}, John D. Hastings\,\orcidlink{0000-0003-0871-3622}\textsuperscript{†}, Shengjie Xu\,\orcidlink{0000-0003-2019-5186}\textsuperscript{‡}, and Adam Fortier\,\orcidlink{0009-0000-1914-7561}\textsuperscript{§}%

\thanks{\textsuperscript{*}The Beacom College of Computer and Cyber Sciences, Dakota State University, Madison, SD, USA. Email: gary.mccully@ieee.org}
\thanks{\textsuperscript{†}The Beacom College of Computer and Cyber Sciences, Dakota State University, Madison, SD, USA. Email: john.hastings@dsu.edu}
\thanks{\textsuperscript{‡}Department of Cyber, Intelligence, and Information Operations, College of Applied Science \& Technology, University of Arizona, Tucson, AZ, USA. Email: sjxu@arizona.edu}
\thanks{\textsuperscript{§}College of Computing, Georgia Institute of Technology, Atlanta, GA, USA. E-mail: afortier8@gatech.edu}}

\maketitle

\begin{abstract}
Detecting vulnerabilities within compiled binaries is challenging due to lost high-level code structures and other factors such as architectural dependencies, compilers, and optimization options. To address these obstacles, this research explores vulnerability detection using natural language processing (NLP) embedding techniques with word2vec, BERT, and RoBERTa to learn semantics from intermediate representation (LLVM IR) code. Long short-term memory (LSTM) neural networks were trained on embeddings from encoders created using approximately 48k LLVM functions from the Juliet dataset. This study is pioneering in its comparison of word2vec models with multiple bidirectional transformers (BERT, RoBERTa) embeddings built using LLVM code to train neural networks to detect vulnerabilities in compiled binaries. Word2vec Skip-Gram models achieved 92\% validation accuracy in detecting vulnerabilities, outperforming word2vec Continuous Bag of Words (CBOW), BERT, and RoBERTa. This suggests that complex contextual embeddings may not provide advantages over simpler word2vec models for this task when a limited number (e.g. 48K) of data samples are used to train the bidirectional transformer-based models. The comparative results provide novel insights into selecting optimal embeddings for learning compiler-independent semantic code representations to advance machine learning detection of vulnerabilities in compiled binaries.
\end{abstract}

\begin{IEEEkeywords}
Machine Learning, Buffer Overflows, BERT, RoBERTa, Binary Security, LLVM, word2vec
\end{IEEEkeywords}

\section{Introduction}
Microsoft is one of the dominant players within the server and desktop operating system market; therefore, vulnerabilities in its software can negatively impact millions of systems around the world. Some of the most prolific worms in the world have exploited vulnerabilities in software developed by Microsoft; these worms include Code Red~\cite{RN72}, which exploited a vulnerability in Microsoft IIS, Sasser~\cite{RN73}, which exploited the Windows LSASS process, and Conficker~\cite{RN74} and WannaCry~\cite{RN75} which exploited SMB vulnerabilities in Windows systems. The financial impact of recovering from these worms is estimated to be billions of dollars (USD)~\cite{RN76}. Microsoft's closed-source nature ~\cite{RN77} prevents organizations from reviewing code written in a high-level programming language for vulnerabilities before deploying it within their environments. Microsoft is a single example of an organization that produces closed-source products but is certainly not the only company. Most commercial organizations protect their intellectual property by not releasing their source code to the public. Therefore, organizations that use these commercial products rely entirely on the organizations that create the software to follow secure coding practices. However, they cannot independently verify the security of the code.

Locating vulnerabilities within compiled binaries can be difficult because significant information is lost when source code is compiled on a target platform. The names of variables, programming structures, and code comments are reduced to low-level machine code. Furthermore, machine code is significantly different for each platform on which it is compiled (e.g., x86/x64, MIPS, SPARC, ARM, etc.). To solve these issues, many researchers have turned to machine learning in the hope that it may provide a solution for locating vulnerabilities in compiled binaries. 

Due to perceived similarities between natural languages and code, researchers investigating machine learning for vulnerability detection in compiled binaries have investigated using natural language processing (NLP) techniques for this purpose. The current body of knowledge extensively utilized neural networks that are commonly used within NLP, including LSTM, gated recurrent unit (GRU), bidirectional LSTM (BLSTM), and bidirectional GRU (BGRU) neural networks. Furthermore, NLP encoding models such as word2vec, instruction2vec, BERT, and RoBERTa have been used to learn semantic relationships within the code. The embeddings these models generate are commonly used to train neural networks to detect vulnerabilities in decompiled code. A summary of this previous work is given in Section ~\ref{related}.

Although researchers have explored the use of specific NLP encoding techniques in the context of decompiled code, research has not yet focused on comparative analysis using the most popular NLP encoding models to identify which embedding techniques perform best in the task of identifying vulnerabilities in compiled binaries. Very few studies have used bidirectional transformer-based models to acquire semantics within decompiled code for this task. RoBERTa was used by \cite{RN67} to learn semantics within the x86 assembly language, and \cite{gallagherllvm} used this model to learn the semantics of the LLVM code. Additionally, \cite{koo2021semantic} uses BERT for code similarity detection, and \cite{han2022binary} trained a BERT model using pcode and used the embeddings to differentiate between vulnerable and non-vulnerable code. Furthermore, according to our research, there have been no comparative studies on the performance of neural networks trained using embeddings from common NLP models (word2vec, BERT, and RoBERTa)  to differentiate between vulnerable and non-vulnerable LLVM code. This paper explores the use of multiple bidirectional encoding models (BERT and RoBERTa) to evaluate their effectiveness in generating embeddings that can be used to accurately detect vulnerabilities in compiled binaries. The models built using these bidirectional embedding models are compared to those generated using word2vec: CBOW and Skip-Gram embeddings. 

This paper can be viewed as an extension of previous work of \cite{RN9} and \cite{schaad2022deeplearningbased}. In these studies, the researchers compiled code samples from the SARD dataset, lifted them to LLVM using the RetDec~\cite{RN22} tool, and used the lifted code to build a word2vec model. Once this model was built, the embeddings from the word2vec model were provided to neural networks to train them to differentiate between vulnerable and non-vulnerable code. However, the research articulated in this paper acknowledges the limitations of word2vec's ability to create context-aware embeddings and explores the alternative use of bidirectional transformer-based models.

The remainder of the paper is organized as follows. Section \ref{design} details the design and implementation, and section \ref{results} presents the results. Section \ref{related} describes the work related to the current research. Section \ref{future} presents ideas for future work, followed by the conclusion in Section \ref{conclusion}.

\section{Design and Implementation}\label{design}
\begin{figure}[!htbp]
    \centering
    \includegraphics[width=1\linewidth]{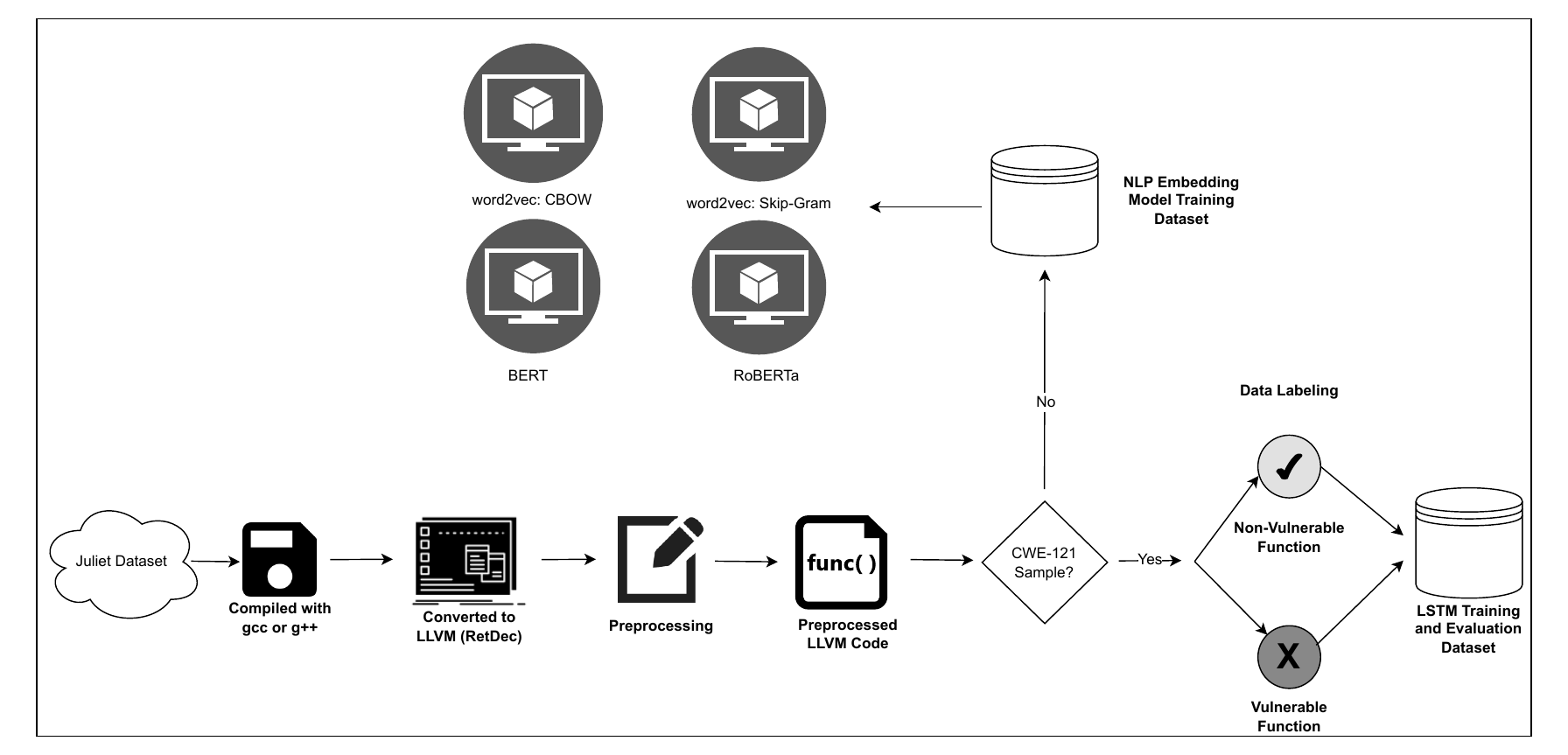}
    \caption{Building Embedding Models Using Lifted Code}
    \label{fig:steps1-4}
\end{figure}

\begin{figure}[!htbp]
    \centering
    \includegraphics[width=1\linewidth]{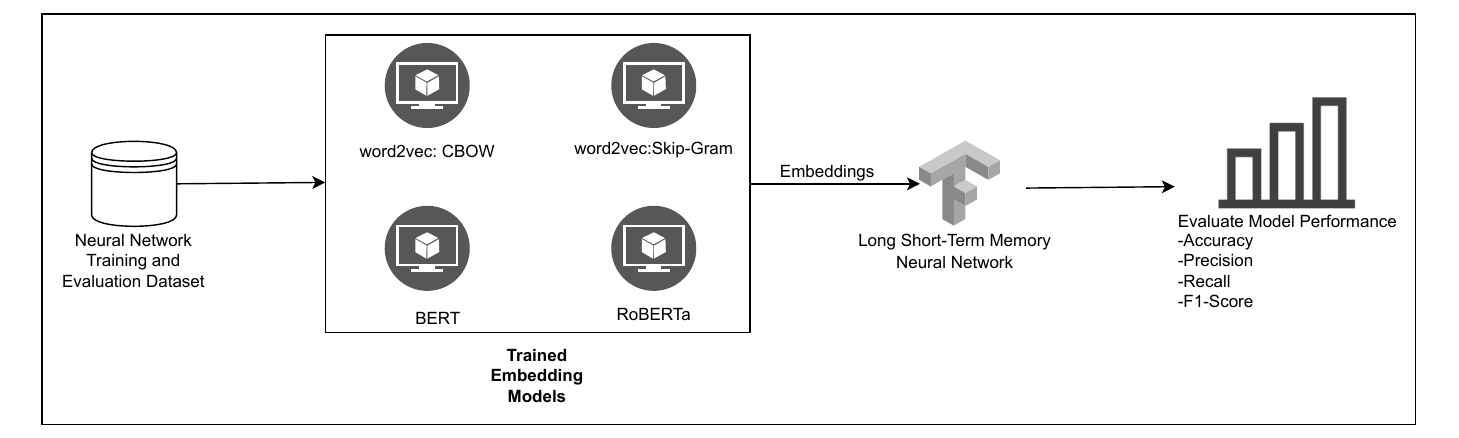}
    \caption{Training Neural Networks to Detect Buffer Overflows}
    \label{fig:steps5-6}
\end{figure}

The current research uses machine learning techniques to identify vulnerabilities in compiled binaries by first using NLP bidirectional encoders to learn code-level semantics in lifted LLVM code. The embeddings generated from these encoders were then used to train LSTM neural networks, and the performance of these neural networks was compared with neural networks trained using simpler word2vec embedding models. An overview of the steps involved in this process is as follows. Fig. \ref{fig:steps1-4} is a visual representation of steps one through four, and \ref{fig:steps5-6} depicts steps five and six of this process.

\begin{enumerate}
\item All code samples, with the exception of samples containing Stack-based Buffer Overflow (CWE-121), in the NIST SARD Juliet~\cite{RN89} dataset were compiled into object files, and the object files were lifted to LLVM. Although the Juliet dataset contains several classes of vulnerabilities and a mix of code samples containing vulnerable and non-vulnerable functions, the primary purpose of the dataset was to learn the semantics between tokenized LLVM instructions. Therefore, tracking vulnerability classes and vulnerable/non-vulnerable status was not required. This process is located in \ref{embeddingdataset}.
\item A second dataset was created that contained vulnerable and non-vulnerable LLVM functions generated using samples containing CWE-121 vulnerabilities taken from the Juliet dataset. The vulnerability status of each of these code samples was tracked within this dataset. The process of creating this data set is described in \ref{nndataset}.
\item The LLVM functions that comprised both datasets were isolated and processed to convert the code into tokens that resemble a natural language. The goal of this phase was to convert the raw LLVM code to a format that could be easily provided to an NLP embedding model. The preprocessing performed is described in \ref{preprocessing}.
\item The processed LLVM code in the first dataset was used to train four NLP embedding models. Two of these embedding models, BERT and RoBERTa, are well-known NLP bidirectional models, and two are much simpler embedding models used extensively throughout the current literature. The creation of the word2vec model is covered in \ref{word2vec}, and the creation of bidirectional encoders is covered in \ref{bidirectionalencoder}.
\begin{itemize}
\item BERT
\item RoBERTa
\item word2vec: CBOW embedding
\item word2vec: Skip-Gram embedding
\end{itemize}
\item The second dataset that comprised both vulnerable and non-vulnerable functions was provided to each embedding model to generate embeddings sent to the LSTM neural network to train the model to detect functions containing CWE-121 vulnerabilities. The LSTM creation process is described in \ref{lstmconfig}.
\item Once each LSTM network was trained, the performance of the model was evaluated using accuracy, precision, recall, and the F1-score. The performance of each LSTM is described in \ref{results}.
\end{enumerate}

\subsection{Embedding Dataset Creation}\label{embeddingdataset}
The dataset chosen for this study is the C/C++ code samples from the Juliet 1.3 dataset within the larger SARD dataset~\cite{RN11}. The rationale for selecting this dataset lies in its use in several other research projects \cite{RN50,RN43,RN67,RN56,RN47,RN46,RN55,schaad2022deeplearningbased}. Choosing a dataset used within multiple research projects provides a basis for a meaningful comparison with the previous corpus of knowledge, as well as for easier reproducibility of results.

The Juliet dataset is comprised of source code samples for several classes of vulnerabilities. For each vulnerability, the dataset contains both a vulnerable and a non-vulnerable version of the source code, and can thus be used to train machine learning models to detect various vulnerabilities. The function names within the code samples differentiate between the vulnerable and non-vulnerable versions.

The Juliet data set, excluding samples containing CWE-121, was compiled to separate object files on a Kali 2023.3 system\footnote[1]{The samples were compiled into object files on a Windows system using both Clang and CL. However, RetDec removed calls to the C Standard Library when lifting the object files to LLVM. As a result, these nonrepresentative samples were excluded from the datasets used to train LLMs and LSTM models.}. Once each source code sample was compiled into an object file, the RetDec tool converted the object files into an LLVM file. Many articles analyzed during the literature review described research on building models to detect vulnerabilities within x86/x64 assembly language instructions. However, this approach limits the applicability of the results to that instruction set, and the same process cannot be applied to additional architectures. Thus, the current research project used RetDec to lift binaries from x86/x64 to an intermediary language called LLVM. The compiler-agnostic~\cite{RN18} nature of LLVM allows for future expansion of datasets for samples compiled using different architectures. Additionally, using LLVM should help reduce some differences between compiler-specific optimizations~\cite{RN18}. Theoretically, a lifted binary for x86/x64 should be identical to a lifted binary compiled for a different architecture (e.g., ARM, SPARK, etc.). 

This first dataset included all functions from the lifted binaries and served the purpose of training embedding models to grasp code-level semantics. The breakdown of the number of functions within this dataset is given in Table~\ref{tab:EmbeddingTrainingDataset}.

\subsection{Neural Network Dataset Creation}\label{nndataset}

Creating the second dataset involved compiling Juliet samples containing CWE-121 into object files and using RetDec to lift these samples to LLVM. Once lifted to LLVM, the clean and vulnerable functions were extracted from the LLVM files. The breakdown of the number of functions within this dataset is in Table ~\ref{tab:EmbeddingTrainingDataset}.

\begin{table}[!htbp]
\caption{Embedding and LSTM Training Datasets}
\label{tab:EmbeddingTrainingDataset}
\center
{%
\begin{tabular}{|l|c|c|}
\hline
& \multicolumn{2}{c|}{\textbf{Training (size)}}  \\ \cline{2-3}
\textbf{Dataset Purpose} & LLM & LSTM \\ \hline

\textbf{Object Files	}& 61,023 & 4,949  \\ \hline

\textbf{LLVM Files}	& 61,023 & 4,949 \\ \hline
\textbf{Extracted Functions:}	& & \\

Pre Duplicate Removal	& 457,529 & \makecell{12,069 \\ \Xhline{0.5pt} Clean: 7,011\\ Vuln: 5,058 } \\ \hline

Post Duplicate Removal	& 48,157 & \makecell{3,901\\ \Xhline{0.5pt} Clean: 2,452\\ Vuln: 1,449} \\ \hline

\makecell[l]{Post Removal of Functions\\ More than 2048 Tokens\\} & NA &
\makecell[c]{ 3,802\\ \Xhline{0.5pt} Clean: 2,386\\ Vuln: 1,416 } \\ \hline
\end{tabular}%
}
\end{table}

Creating the dataset of vulnerable and non-vulnerable functions presented notable challenges due to the naming convention employed by the Juliet dataset. In some cases, the base function (e.g., FunctionName\_Bad) may have a vulnerability embedded within the function itself. However, in other cases, the function representing the vulnerable or non-vulnerable version of the code may be a simple function that calls secondary functions that contain the vulnerability (e.g., Function\_Name\_BadSink). Additionally, in many cases, the non-vulnerable version of the code contained a function that included the keyword Good2Bad or Bad2Good. Thus, when building a dataset representing vulnerable and non-vulnerable versions of the code, all functions containing keywords used to indicate a non-vulnerable version of the code (i.e., Good, goodG2B, goodB2G) and vulnerable versions of the code (i.e., Bad) were included in the dataset. To reduce the number of functions with the sole purpose of calling secondary functions, duplicate versions of each function were removed after the preprocessing articulated in section ~\ref{preprocessing}. 

Tokens from both vulnerable and non-vulnerable functions were fed into NLP embedding models to create embeddings for training LSTM models. Bidirectional encoding models such as BERT and RoBERTa require a max\_length parameter, restricting the number of tokens per processed function. Due to constraints in GPU memory, setting a max\_length significantly beyond 2048 was not feasible. Consequently, samples exceeding the specified max\_length were excluded from the dataset during the training and validation of neural networks.

\subsection{LLVM Preprocessing}\label{preprocessing}
Once the LLVM datasets were generated, a sampling of the LLVM files was manually analyzed to determine the best processing approach. The primary goal of the embedding algorithms was to learn the semantics of the LLVM code. Thus, preprocessing was performed to reduce the number of tokens to focus more on execution order and less on function-specific code offsets. A Jupyter Notebook was created that performed the actions listed in Table ~\ref{tab:preproctabel}.

\begin{table}[!htbp]
\caption{Preprocessing Steps} 
\label{tab:preproctabel}
\resizebox{\columnwidth}{!}{%
\begin{threeparttable}
\begin{tabular}{lcc}
\multicolumn{1}{c}{\textbf{Action}} & \textbf{Before} & \textbf{After} \\
\begin{tabular}[c]{@{}l@{}}LLVM functions \\ created within the \\ code were given a \\ unique identifier\end{tabular} & @\_func\_unique\_name() & func\_x () \\ \hline
\begin{tabular}[c]{@{}l@{}}Function names \\ that were not defined,\\ but were called\\ within the code\\ were given a \\ unique identifier\end{tabular} & @\_undefined\_func\_unique\_name() & extfuncvar\_x () \\ \hline
\begin{tabular}[c]{@{}l@{}}global variables, \\ labels, stack pointers,\\ and numeric metadata\\ were renamed \\ to generic tokens\end{tabular} & \begin{tabular}[c]{@{}c@{}}@global\_var\_offset\\ dec\_label\_pc\_offset:\\ \%stack\_var\_offset\\ ^^21123\end{tabular} & \begin{tabular}[c]{@{}c@{}}globalvar\_x\\  declabel\_x\\     stackvar\_x\\ ^^21x\end{tabular} \\ \hline
\begin{tabular}[c]{@{}l@{}}Digits were isolated \\from the surrounding\\ characters and converted\\ into their word forms\end{tabular} & \begin{tabular}[c]{@{}c@{}}123\end{tabular} & \begin{tabular}[c]{@{}c@{}}one two three \end{tabular} \\ \hline
\begin{tabular}[c]{@{}l@{}}Special characters\\ were renamed to text-\\ based tokens\end{tabular} & \begin{tabular}[c]{@{}c@{}}\{\\    .\\       ,\end{tabular} & \begin{tabular}[c]{@{}c@{}}opcurl\\   dotmark\\    commark\end{tabular}
\end{tabular}%

\begin{tablenotes}
      \small
      \item [a] x is a number used to track the order in which it was defined.
    \end{tablenotes}
  \end{threeparttable}
  
}
\end{table}

\subsection{word2vec: CBOW and Skip-Gram Model Creation}\label{word2vec}
As previously stated, the research team used word2vec: CBOW and Skip-Gram embeddings, BERT, and RoBERTa for generating embeddings. word2vec was selected due to its prevalence in the current body of research~\cite{RN68,RN57,RN36,RN7,RN9,RN58,schaad2022deeplearningbased}, and its ability to provide a strong comparison point with the more complex NLP bidirectional encoders. The techniques used by word2vec -- CBOW and Skip-Gram -- and the code that implements them were released in 2013~\cite{mikolov2013efficient}.

The first of these techniques, CBOW, generates a vector that uniquely represents a particular word using the words before and after. It is especially good at identifying similarities between words because similar words are often found within the same word patterns. For example, in the sentence ``One of my favorite foods to eat is \textit{ice} cream,'' the word \textit{ice} could easily be replaced with \textit{whipped} or even \textit{sour}. Thus, these words would have vectors that highlight that they are similar. However, the word \textit{road} would have a vector that is quite different than \textit{ice}, \textit{whipped}, or \textit{sour} because the words before and after the word are generally quite different from those for the other words. The CBOW technique predicts a missing word using the words surrounding the target word. The second technique used by word2vec, Skip-Gram, attempts to predict surrounding words using a target word, unlike CBOW, which predicts a target word based on the surrounding words.    

The word2vec: CBOW and Skip-Gram models were created using the gensim (4.3.0) Python library. Each preprocessed LLVM function was tokenized using the word tokenizer function of the nltk (3.8.1) library. The model was built using all the 48,157 samples from the embedding model dataset. There were 332 unique tokens used to train these models, and the vector size of the output embedding was set to 100. The window size was set to five, and the minimum count required for each token was set to one. 

\subsection{BERT and RoBERTa Model Creation}\label{bidirectionalencoder}
The code for the BERT model was released in 2019~\cite{devlin2019bert} and stands for Bidirectional Encoder Representations from Transformers. BERT is a bidirectional encoder that learns token embeddings by masking out tokens within the input data and attempting to predict the missing token. Whereas some of the current models at the time were unidirectional, BERT proposed a bidirectional model that scored substantially higher on various NLP tasks than its predecessors. RoBERTa (Robustly Optimized BERT pre-training Approach) was also released in 2019~\cite{liu2019roberta}. RoBERTa is similar to BERT, with some optimizations used to increase its performance.

In this research, the custom BERT model for learning semantics between LLVM instructions was built using the BertForMaskedLM class from Hugging Face's transformers library (4.43.3), while the BertWordPieceTokenizer class from the tokenizers library (0.19.1) was used to tokenize LLVM functions. Similarly, the RoBERTa model was built using the RobertaForMaskedLM and ByteLevelBPETokenizer classes from these libraries. In both cases, 12 hidden layers with 100 nodes were used, each with an output layer of 100. The maximum vocabulary length was 2048, and batch sizes were limited to one due to GPU memory limitations. Training the BERT and RoBERTa models requires a training and validation dataset. Thus, the 48,157 samples from the embedding model dataset were split into two datasets; the training dataset comprised 90\% of the overall dataset, and the validation dataset was 10\% of the dataset.

For both the BERT and RoBERTa embedding models, the loss of training and validation steps was shown every 1,000 steps. The best loss in the validation set for the BERT and RoBERTa models was achieved after step 54,000, as shown in Table \ref{tab:Bertrobertaloss}. These were the embedding models used to create the embeddings provided to the LSTM models.

\definecolor{lgreen}{rgb}{0.78, 0.99, 0.78}
\begin{table}[!htbp]
\caption{BERT/RoBERTa Five Best Loss Scores by Validation Step}
\label{tab:Bertrobertaloss}
\center
\resizebox{0.8\columnwidth}{!}{%
\begin{tabular}{ccc|cc}
   & \multicolumn{2}{c}{\textbf{BERT}} & \multicolumn{2}{c}{\textbf{RoBERTa}} \\
\textbf{Step} & Training & Validation & Training & Validation\\
50,000 & 2.555300 & 2.496149 & 2.558400 & 2.494601 \\
51,000 & 2.555500 & 2.497498 & 2.557400 & 2.488815 \\
52,000 & 2.551600 & 2.493027 & 2.559900 & 2.485950 \\
53,000 & 2.546700 & 2.495847 & 2.548000 & 2.487699 \\
54,000 & \cellcolor{lgreen}2.554400 & \cellcolor{lgreen}2.492988 & \cellcolor{lgreen}2.557500 & \cellcolor{lgreen}2.483047
\end{tabular}%
}
\end{table}

\subsection{LSTM Neural Network Configuration}\label{lstmconfig}
This research uses embeddings generated from each encoding model to train LSTM neural networks to identify vulnerabilities within the compiled code. LSTM neural networks are recurrent neural networks designed to avoid vanishing gradients by using gates to track relevant information and forget irrelevant details. One of the advantages of recurrent neural networks is that they take the output from one step and reuse the output in subsequent steps. For example, suppose that an LSTM tries to predict the next word in the sequence, ``my favorite food is ice...'' There is a high probability that the next word is \textit{cream}, so the LSTM would likely predict \textit{cream} as the next word. However, if the next word entered is \textit{cold}; the sentence has changed to ``my favorite food is ice cold...'' The LSTM can use the output of the previous step to update its prediction from \textit{cream} to a different word, such as \textit{custard}. The recurrent nature of the LSTM should make it better suited for learning the patterns within the embeddings needed to identify vulnerable and non-vulnerable code. Although more complex neural networks may result in better performance, an LSTM was chosen because it is a good baseline neural network to evaluate the performance of different embedding models.

For the current research project, the LSTM neural networks comprised two hidden layers with 128 neurons each and 50 epochs to train each network. The Leaky Rectified Linear Unit (LeakyReLU) activation function was chosen because recurrent neural networks are prone to the problem of vanishing gradients~\cite[p.~335]{RN25}, and the LeakyReLU activation function helps mitigate this problem. Each layer of the LSTM model was created using the TensorFlow and Keras libraries. A dropout rate of 20\% was used between the LSTM layers, and a single dense layer of a single neuron was used in the output layer. The dataset was randomized and divided into two subsets. The first subset was used to train each neural network and consisted of 80\% (i.e., 3,041 samples) of the overall dataset; the second subset was used for model validation and consisted of 20\% (i.e., 761) of the total dataset. The same training and validation subsets were used for each neural network. 

All neural networks were trained twice: once with the hidden layers of the embedding model frozen and once with them unfrozen. Each LSTM was created using several different hyperparameters to better understand the effects of optimizers and learning rates on model performance. The hyperparameters used for training each neural network are as follows:
\begin{itemize}
\item Adam optimizer: learning rates of 0.01, 0.001, and 0.0001
\item SGD optimizer: a learning rate and momentum of 0.01
\item SGD: a learning rate of 0.0001 and momentums of 0.01, 0.001, and 0.0001.
\end{itemize}

\section{Results}\label{results}
Accuracy, precision, recall, and F1-score were evaluated for each LSTM neural network created from the embeddings generated from each embedding model. Although each neural network was trained with and without freezing the embedding model's hidden layers, only the results from the unfrozen models are included due to their superior performance. The metric tables in the following section do not include models with hyperparameters that did not achieve accuracy gains beyond the base accuracy of placing all samples in a single category.

\subsection{word2vec: Skip-Gram Metrics}
As shown in Table \ref{tab:skipPerf}, the model with the highest performance using word2vec Skip-Gram embeddings achieved an accuracy of 92.0\%, making it the model with the best performance in this study. This model was generated using the Adam optimizer with a learning rate of 0.001. The SGD optimizer results with a learning rate of 0.0001 and a momentum of 0.01, 0.001, and 0.0001 struggled to achieve accuracy gains beyond the base accuracy of placing all samples in a single category. However, these models consistently achieved lower loss values for each epoch, which indicates that additional epochs may eventually increase model performance.

\begin{table}[!htbp]
\caption{word2vec: Skip-Gram Metrics}
\label{tab:skipPerf}
\resizebox{\columnwidth}{!}{%
\begin{tabular}{lcccccc}
\makecell[l]{\textbf{Hyper-}\\ \textbf{parameters}} & \textbf{Epoch} & \textbf{Loss} & \textbf{Accuracy} & \textbf{Precision} & \textbf{Recall} &\makecell[l]{\textbf{F1-} \\ \textbf{Score}} \\
\makecell[l]{SGD\\ -LR: 0.01\\ -Mom: 0.01} & 45 & 0.3663 & 80.9\% & 68.2\% & 90.0\% & 77.6\% \\ \hline
\makecell[l]{Adam\\ -LR: 0.01} & 35 & 0.4127 & 81.1\% & 69.1\% & 87.5\% & 77.2\% \\ \hline
\makecell[l]{Adam\\ -LR: 0.001} & \cellcolor{lgreen}43 & \cellcolor{lgreen}0.1848 & \cellcolor{lgreen}92.0\% & \cellcolor{lgreen}85.2\% & \cellcolor{lgreen}94.6\% & \cellcolor{lgreen}89.6\% \\ \hline
\makecell[l]{Adam\\ -LR: 0.0001} & 48 & 0.3282 & 84.0\% & 72.1\% & 91.8\% & 80.8\% \\ \hline
\end{tabular}%
}
\end{table}

\subsection{word2vec: CBOW Metrics}
The metrics in Table ~\ref{tab:cbowPerf} were recorded when each model reached its highest accuracy. The highest accuracy, 87.5\%, was obtained when the Adam optimizer was used with a learning rate of 0.001. LSTMs trained with the SGD optimizer and a low learning rate exhibited poor performance. However, similar to models trained with Skip-Gram embeddings, consistent improvements in loss scores across epochs suggest that additional training could have further enhanced performance.

\begin{table}[!htbp]
\caption{word2vec: CBOW Metrics}
\label{tab:cbowPerf}
\resizebox{\columnwidth}{!}{%
\begin{tabular}{lcccccc}
\makecell[l]{\textbf{Hyper-}\\ \textbf{parameters}} & \textbf{Epoch} & \textbf{Loss} & \textbf{Accuracy} & \textbf{Precision} & \textbf{Recall} &\makecell[l]{\textbf{F1-} \\ \textbf{Score}} \\
\makecell[l]{SGD\\ -LR: 0.01\\ -Mom: 0.01} & 43 & 0.3417 & 83.0\% & 71.8\% & 88.5\% & 79.3\% \\ \hline
\makecell[l]{SGD\\ -LR: 0.0001\\ -Mom: 0.01} & 50 & 0.5920 & 69.8\% & 67.9\% & 33.3\% & 44.7\% \\ \hline
\makecell[l]{SGD\\ -LR: 0.0001\\ -Mom: 0.001} & 42 & 0.6246 & 69.0\% & 64.2\% & 34.8\% & 45.1\% \\ \hline
\makecell[l]{SGD\\ -LR: 0.0001\\ -Mom: 0.0001} & 37 & 0.6159 & 68.7\% & 63.4\% & 34.8\% & 44.9\% \\ \hline
\makecell[l]{Adam\\ -LR: 0.01} & 46 & 0.3531 & 83.4\% & 70.7\% & 93.5\% & 80.6\% \\ \hline
\makecell[l]{Adam\\ -LR: 0.001} & \cellcolor{lgreen}34 & \cellcolor{lgreen}0.2528 & \cellcolor{lgreen}87.5\% & \cellcolor{lgreen}78.0\% & \cellcolor{lgreen}91.8\% & \cellcolor{lgreen}84.3\% \\ \hline
\makecell[l]{Adam\\ -LR: 0.0001} & 45 & 0.2990 & 86.5\% & 77.0\% & 90.0\% & 83.0\% \\ \hline
\end{tabular}%
}
\end{table}

\subsection{RoBERTa Metrics}
The LSTM model with the best performance achieved an accuracy of 88.8\% and an F1-score of 84.2\% using the SGD optimizer with a learning rate of 0.0001 and a momentum of 0.001. Table ~\ref{tab:RoBERTaPerf} shows that most of the models trained with the SGD optimizer achieved an accuracy of approximately 88\%. However, when using the Adam optimizer, these models did not achieve meaningful accuracy gains.

\begin{table}[!htbp]
\caption{RoBERTa Metrics}
\label{tab:RoBERTaPerf}
\resizebox{\columnwidth}{!}{%
\begin{tabular}{lcccccc}
\makecell[l]{\textbf{Hyper-}\\ \textbf{parameters}} & \textbf{Epoch} & \textbf{Loss} & \textbf{Accuracy} & \textbf{Precision} & \textbf{Recall} &\makecell[l]{\textbf{F1-} \\ \textbf{Score}} \\
\makecell[l]{SGD\\ -LR: 0.01\\ -Mom: 0.01} & 31 & 0.2625 & 86.6\% & 82.2\% & 81.0\% & 81.6\% \\ \hline
\makecell[l]{SGD\\ -LR: 0.0001\\ -Mom: 0.01} & 47 & 0.2454 & 88.7\% & 79.0\% & 94.3\% & 85.9\% \\ \hline
\makecell[l]{SGD\\ -LR: 0.0001\\ -Mom: 0.001} & \cellcolor{lgreen}45 & \cellcolor{lgreen}0.2708 & \cellcolor{lgreen}88.8\% & \cellcolor{lgreen}87.6\% & \cellcolor{lgreen}81.0\% & \cellcolor{lgreen}84.2\% \\ \hline
\makecell[l]{SGD\\ -LR: 0.0001\\ -Mom: 0.0001} & 49 & 0.2868 & 88.6\% & 84.0\% & 84.9\% & 84.5\% \\ \hline
\end{tabular}%
}
\end{table}

\subsection{BERT Metrics}
The LSTM model that performed best with BERT embeddings was built using the SGD optimizer with a learning rate of 0.0001 and a momentum of 0.001. Table ~\ref{tab:BERTPerf} shows that the LSTM models trained using BERT embeddings and the SGD optimizer achieved a maximum accuracy range of 87-88\%, with the best-performing model reaching an accuracy of 88.8\%. Similarly to models built with RoBERTa embeddings, those using the Adam optimizer struggled to achieve any significant accuracy gains.

\begin{table}[!htbp]
\caption{BERT Metrics}
\label{tab:BERTPerf}
\resizebox{\columnwidth}{!}{%
\begin{tabular}{lcccccc}
\makecell[l]{\textbf{Hyper-}\\ \textbf{parameters}} & \textbf{Epoch} & \textbf{Loss} & \textbf{Accuracy} & \textbf{Precision} & \textbf{Recall} &\makecell[l]{\textbf{F1-} \\ \textbf{Score}} \\
\makecell[l]{SGD\\ -LR: 0.01\\ -Mom: 0.01} & 19 & 0.2862 & 87.5\% & 81.7\% & 84.9\% & 83.3\% \\ \hline
\makecell[l]{SGD\\ -LR: 0.0001\\ -Mom: 0.01} & 20 & 0.2913 & 88.6\% & 81.0\% & 90.0\% & 85.2\% \\ \hline
\makecell[l]{SGD\\ -LR: 0.0001\\ -Mom: 0.001} & \cellcolor{lgreen}29 & \cellcolor{lgreen}0.2625 & \cellcolor{lgreen}88.8\% & \cellcolor{lgreen}80.1\% & \cellcolor{lgreen}92.5\% & \cellcolor{lgreen}85.9\% \\ \hline
\makecell[l]{SGD\\ -LR: 0.0001\\ -Mom: 0.0001} & 47 & 0.2909 & 87.3\% & 83.0\% & 82.1\% & 82.5\% \\ \hline
\end{tabular}%
}
\end{table}

\section{Discussion}\label{discussion}
As shown in Table \ref{tab:modelComp}, neural networks built with word2vec Skip-Gram embeddings achieved the highest performance, reaching a peak accuracy of 92.0\% when using the Adam optimizer with a learning rate of 0.001. Similarly, models utilizing word2vec: CBOW embeddings also demonstrated relatively strong performance, achieving a high accuracy of 87.5\% with the same optimizer and learning rate. However, neither word2vec: CBOW and Skip-Gram performed well when the SGD optimizer was used with a lower learning rate (0.0001). Nevertheless, the behavior displayed by the loss graphs of both word2vec models indicates that additional epochs may have resulted in better performance.

\begin{table}[!htbp]
\caption{Overall LSTM Model Comparison}
\label{tab:modelComp}
\resizebox{\columnwidth}{!}{%
\begin{tabular}{lcccccc}
\makecell[l]{\textbf{Hyper-}\\ \textbf{parameters}} & \textbf{Epoch} & \textbf{Loss} & \textbf{Accuracy} & \textbf{Precision} & \textbf{Recall} &\makecell[l]{\textbf{F1-} \\ \textbf{Score}} \\
\makecell[l]{word2vec\\ Skip-Gram\\ Adam\\ -LR: 0.001} & \cellcolor{lgreen}43 & \cellcolor{lgreen}0.1848 & \cellcolor{lgreen}92.0\% & \cellcolor{lgreen}85.2\% & \cellcolor{lgreen}94.6\% & \cellcolor{lgreen}89.6\% \\ \hline
\makecell[l]{BERT\\ SGD\\ -LR: 0.0001\\ -Mom: 0.001}& 29 & 0.2625 & 88.8\% & 80.1\% & 92.5\% & 85.9\% \\ \hline
\makecell[l]{RoBERTa\\ SGD\\ -LR: 0.0001\\ -Mom: 0.001} & 45 & 0.2708 & 88.8\% & 87.6\% & 81.0\% & 84.2\% \\ \hline
\makecell[l]{word2vec\\ CBOW\\ Adam\\ -LR: 0.001} & 34 & 0.2528 & 87.5\% & 78.0\% & 91.8\% & 84.3\% \\ \hline
\end{tabular}%
}
\end{table}

The higher performance of word2vec Skip-Gram over CBOW is somewhat interesting. After preprocessing, the number of unique tokens within the LLVM functions was limited, with all functions represented by only 332 unique tokens. This was due, in part, to the way that preprocessing limited unnecessary uniqueness within tokenized LLVM functions. The CBOW model's technique tends to produce more accurate embeddings for representing tokens with higher frequency than the Skip-Gram model~\cite{mikolov2013efficient}. However, Skip-Gram models typically produce more accurate embeddings than CBOW models when trained with smaller datasets~\cite{mikolov2013efficient}. Thus, in this research, Skip-Grams' capabilities to represent token-level relationships using datasets of limited size resulted in better performance despite CBOW's accuracy gains achieved through token frequency.

The top accuracy of neural networks trained using BERT and RoBERTa embeddings was lower than the model trained using word2vec Skip-Gram embeddings. This implies that, for smaller datasets, the simpler techniques used by word2vec Skip-Gram outperform the more complex context-aware techniques used by BERT and RoBERTa. This result is not overly surprising considering that the original BERT model was pre-trained using BooksCorpus (800M words) and English Wikipedia (2,500M words)~\cite{devlin2019bert} and RoBERTa was trained on ten times that amount of data~\cite{liu2019roberta}. Thus, it is suspected that neural networks built using embeddings from BERT and RoBERTa models trained on larger datasets would likely achieve higher performance. However, despite the smaller training dataset used to train the BERT and RoBERTa models, it is interesting to note that neural networks utilizing these embeddings, optimized with SGD and a learning rate of 0.0001, outperformed the word2vec models trained under the same conditions. Furthermore, these models achieved relatively strong top performance, achieving high accuracy rates of 88.8\% with RoBERTa and BERT. Thus, even though these models are meant to be trained with large amounts of data, they still manage to perform moderately well with smaller datasets.  

\section{Related Work}\label{related}
The datasets integrated into the National Institute of Science and Technology's (NIST) Software Assurance Reference Dataset \cite{RN11} are among the most frequently referenced source code repositories in the reviewed literature. \cite{RN9,RN10,RN57,RN45} explicitly stated that they used the NIST SARD dataset when selecting samples used for their research. One of the SARD subdatasets, the Juliet dataset, was used by \cite{RN50,RN43,RN67,RN56,RN47,RN46,RN55,schaad2022deeplearningbased}. Two additional subdatasets within SARD used by researchers include MIT Lincoln Laboratories (MLL05b), used by \cite{RN41}, and STONESOUP (IARPA12, IARPA14), used by \cite{RN43}.

Training auxiliary models for learning semantical relationships in code is commonly used throughout the literature. \cite{RN68,RN57,RN36,RN7,RN9,RN58,schaad2022deeplearningbased} used word2vec to generate embedding models to capture semantics within the code. However, due to the limitations of relying on word2vec to represent relationships within decompiled code, other studies propose alternative embedding models similar to word2vec or built upon word2vec embeddings. These alternatives include Instruction2vec~\cite{RN43}, Bin2vec~\cite{RN47}, VDGraph2Vec~\cite{RN67}, BinVulDet~\cite{RN46}, and VulANalyzeR~\cite{RN55}). 

Few papers used a common NLP embedding model other than word2vec in their research. \cite{RN67} used RoBERTa within their process to learn semantic relationships within assembly code and passed these embeddings to a Message Passing Neural Network (MPNN) for vulnerability detection. \cite{gallagherllvm} custom-built a RoBERTa model with LLVM code to learn the code-level semantics to detect vulnerable code. \cite{han2022binary} trained a BERT model with pcode and utilized the generated embeddings to train machine learning models to differentiate between vulnerable and non-vulnerable code. Finally, \cite{koo2021semantic} used BERT, and \cite{10.1145/3460120.3484587} used a GPT model to create embeddings for code similarity detection. 

Recurrent neural networks commonly used in NLP are regularly used to detect vulnerabilities in compiled binaries. For example, \cite{RN36,RN40,RN10,RN9,RN57,RN47,RN51,RN58,schaad2022deeplearningbased} used an LSTM neural network and \cite{RN9,RN42,RN57,schaad2022deeplearningbased} applied a GRU neural network for vulnerability detection. Furthermore, \cite{RN7,RN10,RN9,RN57,RN45,RN51,RN55,schaad2022deeplearningbased} used a BLSTM, and \cite{RN10,RN9,RN57,RN45,RN46,RN51,RN55,schaad2022deeplearningbased} used a BGRU neural network as part of the research they performed.

The papers most similar to the experiment performed within the current paper are \cite{RN9} and \cite{schaad2022deeplearningbased}. Both papers lift assembly language to LLVM using RetDec, perform preprocessing on LLVM, use word2vec to create embeddings, and use a neural network to train a neural network to discover vulnerabilities in the LLVM code. This paper builds on their research by expanding the study to include additional embedding models (BERT and RoBERTa).

\section{Future Research}\label{future}
It is postulated that the BERT and RoBERTa embeddings may perform better with more data, so a significant emphasis in future research will be on identifying ways to increase the number of quality samples used to train the embedding models. Furthermore, it is possible that additional vulnerable and non-vulnerable code samples could also affect LSTM neural network performance. Thus, a second emphasis for future research would be identifying ways to increase the number of code samples that could be used to train these models.  

Future research will increase its focus on unidirectional embedding models, including GPT-1 and GPT-2~\cite{mccully2024}. A different NLP embedding model may achieve better results when there are fewer samples for training embedding models. Furthermore, some researchers~\cite{RN7} were able to achieve accuracy gains when adding bidirectionally to their neural networks. Thus, future research will evaluate different neural networks to identify the best performance. Finally, the current study strictly focused on stack-based buffer overflow vulnerabilities. However, it would make sense to expand the research to include additional vulnerability categories in the future. 

Finally, the study \cite{arp2022and} states that data snooping is one of the biggest pitfalls in current research on the application of machine learning to cyber security. Data snooping occurs when training and validation data are not properly separated. The current study does not suffer from data snooping, since the samples within the dataset used to train the LSTM models were never used to train any of the LLM models. However, due to the prominence of data snooping in the current literature~\cite{arp2022and}, it would be worthwhile to perform additional experiments to understand just how significant the effect of data snooping would be on current results. 

\section{Conclusion}\label{conclusion}
The most significant contribution of the current research is that it is the first study to explore the use of multiple bidirectional transformers (i.e. BERT and RoBERTa) embeddings built using LLVM code to train neural networks to detect vulnerabilities in the lifted code. In addition, it explored the effects of Adam and SGD optimizers, comparing the performance of models using different values for learning rates and momentum. This study also identified vital limitations that more complex NLP embeddings introduce to the process of using embeddings for vulnerability discovery in compiled binaries, in contrast to simpler word2vec embeddings. For example, increased embedding max\_lengths results in significant memory consumption that limits the maximum number of tokens that can be processed in any code block. 

At the end of the project, it was determined that, when using embeddings generated from NLP models trained using a limited number of samples (i.e., 48K), the LSTM built using word2vec: Skip-Gram embeddings with the Adam optimizer and an initial learning rate of 0.001 achieved the highest accuracy (92.0\%). However, this top accuracy is only slightly better than those of RoBERTa and BERT (88.8\%). The results suggest that complex contextual NLP embeddings may not provide advantages over simpler word2vec models for this task when these models are trained using smaller datasets. The comparative results provide novel insights into selecting optimal embeddings for learning compiler-independent semantic code representations to advance machine learning detection of vulnerabilities in compiled code.

\balance
\printbibliography

\end{document}